# Deposition distribution of the new coronavirus (SARS-CoV-2) in the human airways upon exposure to cough-generated aerosol

**Balázs G. Madas[1*], Péter Füri[1], Árpád Farkas[1], Attila Nagy[2], Aladár Czitrovszky[2], Imre Balásházy[1], Gusztáv G. Schay[3], Alpár Horváth[4,5]**

[1]Environmental Physics Department, Centre for Energy Research, Budapest, Hungary

[2]Laser Applications and Optical Measurement Techniques, Applied and Nonlinear Optics, Institute for Solid State Physics and Optics, Wigner Research Centre for Physics, Budapest, Hungary

[3]Department of Biophysics and Radiation Biology, Semmelweis University, Budapest, Hungary

[4]Department of Pulmonology, County Institute of Pulmonology, Törökbálint, Hungary

[5]Medical Department, Chiesi Hungary Ltd, Budapest, Hungary

**\* Correspondence:**
Corresponding Author
balazs.madas@energia.mta.hu

## Abstract

The new coronavirus disease 2019 (COVID-19) has been emerged as a rapidly spreading pandemic. The disease is thought to spread mainly from person-to-person through respiratory droplets produced when an infected person coughs, sneezes, or talks. The pathogen of COVID-19 is the severe acute respiratory syndrome coronavirus 2 (SARS-CoV-2). It infects the cells binding to the angiotensin-converting enzyme 2 receptor (ACE2) which is expressed by cells throughout the airways as targets for cellular entry. Although the majority of persons infected with SARS-CoV-2 experience symptoms of mild upper respiratory tract infection, in some people infections of the peripheral airways result in severe, potentially fatal pneumonia. However, the induction of COVID-19 pneumonia requires that SARS-CoV-2 reaches the peripheral airways. While huge efforts have been made to understand the spread of the disease as well as the pathogenesis following cellular entry, much less attention is paid how SARS-CoV-2 from the environment reach the receptors of the target cells. The aim of the present study is to characterize the deposition distribution of SARS-CoV-2 in the airways upon exposure to cough-generated aerosol. For this purpose, the Stochastic Lung Deposition Model has been applied. Aerosol size distribution and breathing parameters were taken from the literature supposing normal breathing through the nose. We found that the probability of direct infection of the peripheral airways due to inhalation of aerosol generated by a bystander cough is very low. As the number of pathogens deposited in the extrathoracic airways is ~10 times higher than in the peripheral airways, we concluded that in most cases COVID-19 pneumonia must be preceded by SARS-CoV-2 infection of the upper airways. Our results suggest that without the enhancement of viral load in the upper airways, COVID-19 would be much less dangerous. The period between the onset of initial symptoms and the potential clinical deterioration could provide an opportunity for prevention of pneumonia by blocking or significantly reducing the transport of viruses towards the peripheral airways. Coughing into a tissue or cloth even at home in order to absorb the emitted aerosol is highly recommended to avoid the continuous re-inhalation of own cough.

# 1   Introduction

The new coronavirus disease 2019 (COVID 19) has been emerged as a rapidly spreading pandemic (1) originating from Wuhan, China (2). There are currently few studies that define the pathophysiological characteristics of COVID-19, and there is great uncertainty regarding its mechanism of spread (3). However, the disease is thought to spread i) mainly from person-to-person, who are in close contact with one another (within about 2 m) ii) through respiratory droplets produced when an infected person coughs, sneezes or talks iii) which can land in the mouths or noses of people who are nearby or possibly be inhaled into the lungs (4). Virological assessment of COVID-19 also suggests that the transmission is droplet-, rather than fomite-, based (5).

The pathogen of COVID 19 is the severe acute respiratory syndrome coronavirus 2 (SARS-CoV-2) (6), which infects the cells binding to the angiotensin-converting enzyme 2 receptor (ACE2) (7) primarily in the respiratory system. Cells expressing ACE2 can be found throughout the airways (8), and therefore, cellular entry of SARS-CoV-2 can also take place throughout the airways (9). Although the majority of persons infected with SARS-CoV-2 experience symptoms of mild upper respiratory tract infection, in some people infections of the peripheral airways result in severe pneumonia potentially leading to significant hypoxia with acute respiratory distress syndrome (ARDS) and death. However, the induction of COVID-19 pneumonia and ARDS requires that SARS-CoV-2 reaches the lower airways.

While huge efforts have been made to understand the spread of the disease as well as the pathogenesis following cellular entry of SARS-CoV-2, much less attention is paid how viruses from the environment reach the receptors of the target cells in the respiratory system. The aim of the present study is to characterize the deposition distribution of pathogens in the airways upon exposure to cough-generated aerosol and discuss its consequences on the pathogenesis of the disease.

# 2   Methods

For this purpose, the most recent version of the Stochastic Lung Deposition Model has been applied (10). It was originally developed by Koblinger and Hofmann (11) and continuously extended during the last three decades (12,13). In this model, the geometry of the airways along the path of an inhaled particle is selected randomly based on statistical analysis of large anatomical databases (14,15), while deposition probabilities are computed by deterministic formulae considering inertial impaction, gravitational settling and Brownian diffusion. The deposition probability in the extrathoracic airways is determined by an empirical deposition formula (16). More details on the Stochastic Lung Deposition Model can be found here (10–13). The deposition model was previously validated against *in vivo* airway deposition measurements.

The model computes the fraction of inhaled particles that deposit in each anatomical region of the lungs. In addition, it also yields the deposition fraction as a function of airway generation number[1]. The fraction of inhaled mass in different anatomical regions and airway generations can also be obtained, which is particularly useful if the pathogen concentration in the coughed material is supposed to be independent on the particle size. These fractions are quantified for a full breathing cycle supposing normal breath through the nose.

---

[1] The first airway generation consists of the trachea and the first half the of the main bronchi. The second airway generation consists of the second half of the main bronchi and the first half of their daughters, and so on.

Functional residual capacity of 3300 cm³, tidal volume of 750 cm³, and breathing frequency of 12 min⁻¹ were taken from the Human Respiratory Tract Model of the International Commission on Radiological Protection corresponding to an adult man in sitting position (17) and applicable for normal breath if the spine is in vertical position. It was supposed that the duration of inhalation is 1.9 s followed by 0.1 s breath hold, and exhalation lasts for 2 s followed by a 1-second-long breath hold.

Besides the geometry and the flow conditions in the airways, the lung deposition of the aerosol particles is determined by their aerodynamic properties which depends on many parameters (e.g. size, shape, morphology, density etc.). Lindsey at al. measured the number size distributions of cough-generated aerosol particles by means of a combined wide range aerosol particle spectrometer (optical particle counter and scanning mobility particle sizer system), when patients with influenza cough (18). Although it is a different disease, the virus size as well as the mechanism of aerosol generation during dry cough is similar in case of COVID-19 and influenza, so we used these data. The mass size distribution was obtained from the concentration, number size distribution and density of cough-generated particles. For the calculations we assumed, that the particles are spherical and their physical size is equal to their measured optical size. The mass size distribution is plotted in Figure 1.

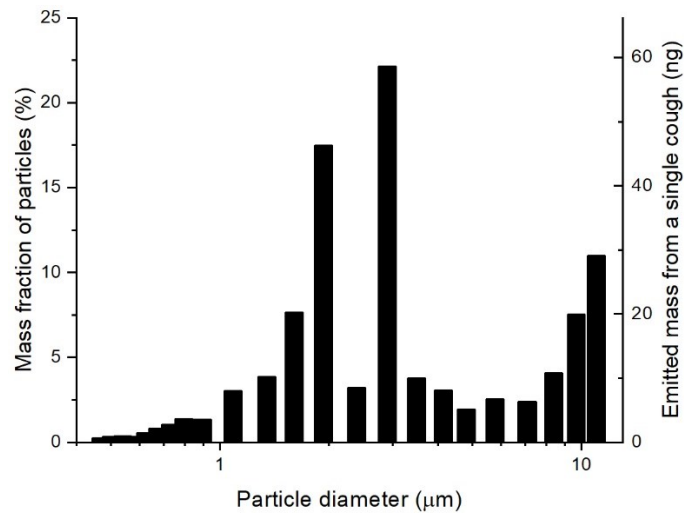

Figure 1. Mass size distribution of particles emitted by coughing of patients with influenza (18). This mass size distribution was used as input in the Monte-Carlo simulations.

The cough-generated particles may travel great distance if the meteorological conditions (air velocity, temperature, humidity) are favorable (19). During the flight the particles may evaporate or fall down by gravitational settling. Therefore, the present simulations apply only to one of supposed major transmission routes, when a bystander subject (within a 1-m-distance) directly inhales the cough-generated aerosol (4). If the inhalation takes place more distant from the source in space or time, then the size distribution has to be measured near to the subject inhaling the emitted particles.

In a recent study, it was found that virus concentration varies highly in throat swab and sputum samples of SARS-CoV-2-infected patients with a maximum concentration exceeding 10¹¹ RNA copies per cm³, while the median values are in the order of 10⁵ - 10⁶ copies per cm³ (20). Values along this wide range were used to obtain upper estimate for pathogen number in different lung regions and airway generations. Other studies provided similar range for viral loads (5,21). It is important to note however that the number of viruses capable of infection is much lower than the number of RNA copies. In a recent study, viruses could not be isolated from samples that contained less than $10^6$ RNA copies per $cm^3$ (5).

We supposed that the virus concentration is independent on the particle size. It means that the number of pathogens is directly proportional to the mass of the particles. It is an appropriate approximation in locations close to the source only, because evaporation increases the virus concentrations in aerosol particles, and evaporation rate strongly depends on the particle size.

## 3    Results

As the infection with SARS-CoV-2 causes very different symptoms depending on the infected region, first we focus on the regional deposition distribution. Figure 2 shows that while 61.8% of the inhaled mass is filtered out by the upper airways, significant fractions deposit in the bronchial (~5.5%) and acinar airways (~8.5%). Considering the average particle number concentration in the material coughed (29,600 $dm^{-3}$), the tidal volume (0.75 $dm^3$), and the mean particle mass (4.6 pg), it can be calculated that a single inhalation results in 63 ng material depositing in the extrathoracic airways, while 5.6 ng and 8.6 ng deposit in the bronchial and acinar airways, respectively.

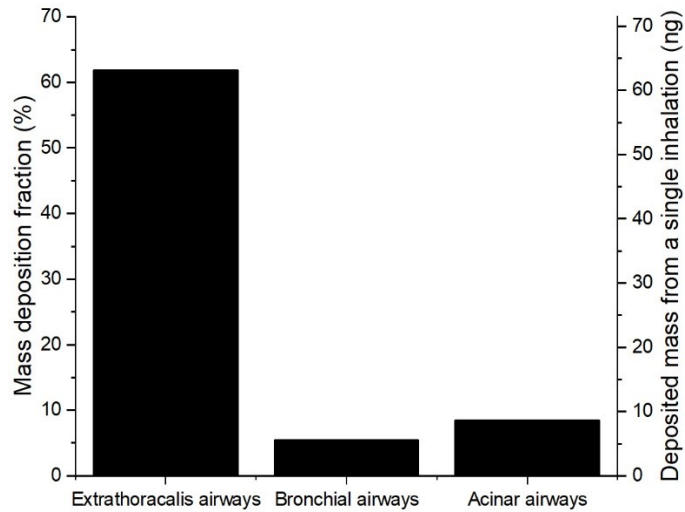

Figure 2. Mass deposition fraction of inhaled particles (left axis) and deposited mass from a single inhalation (right axis) in the extrathoracic, bronchial and acinar regions of the lungs.

In a subtler subdivision of the intrathoracic airways, one can distinguish peripheral airways (all acinar airways except the first four generations, i.e. the bronchiolus respiratorius region), small airways (with a diameter smaller than 2 mm except the peripheral airways), and large airways (with a diameter larger than 2 mm). Figure 3 shows the mass deposition fractions and deposited mass from a single inhalation in these airways. While 2.7 ng and 5.7 ng material deposit in the large and small airways, respectively, 5.8 ng reach the peripheral airways, and deposit there. It may be also of interest that 2.9 ng material deposits in the bronchiolus respiratorius region.

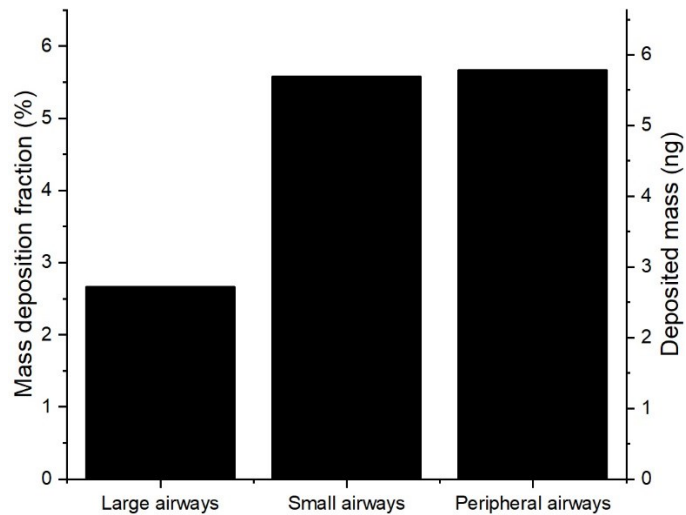

Figure 3. Mass deposition fraction of inhaled particles (left axis) and deposited mass from a single inhalation (right axis) in the large, small, and peripheral airways.

The Stochastic Lung Deposition Model is able to determine the deposition fraction and deposited mass as the function of airway generation number. Figure 4 shows that in terms of deposited mass the most affected part of the acinar airways is the 19[th] and 20[th] airway generations, which means that most of the pathogens penetrating the extrathoracic airways will pass 19 bifurcations before depositing. In the bronchial region, the highest amount deposit in the 12[th] airway generation. The acinar peak is more than three-fold higher than the bronchial one. However, the difference in deposition density is much smaller as the surface of the airways strongly increases with the generation number.

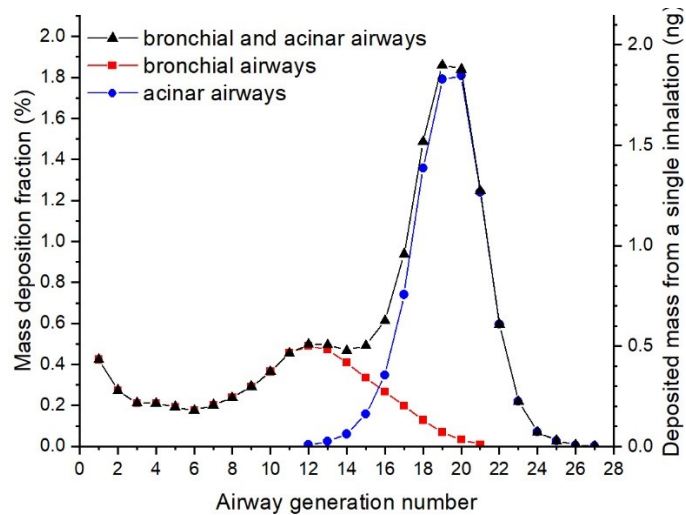

Figure 4. Mass deposition fraction of inhaled particles (left axis) and deposited mass from a single inhalation (right axis) as the function of airway generation number.

In order to estimate the amount of deposited pathogens in different parts of the airways, viral RNA concentrations measured by Pan et al. (20) was used. They found that the viral loads ranged from 641 RNA copy per cm$^3$ to $1.34 \times 10^{11}$ RNA copy per cm$^3$ with a median of $7.99 \times 10^4$ and $7.52 \times 10^5$ RNA copy per cm$^3$ in throat swab and sputum samples, respectively. In order to obtain an upper estimate for the pathogen concentration in the material emitted by dry coughing, the concentrations in the throat have to be taken into account. Therefore, the former of the median values is considered. Using the density from Lindsey et al. (18), these data can be converted to virus concentrations in unit mass resulting in viral loads ranging from 376 RNA copy per g to $7.86 \times 10^{10}$ RNA copy per g with a median of $4.69 \times 10^4$ RNA copy per g in throat swab samples.

Combining the median value in sputum samples with the regional deposition distribution data, it can be found that about 4000 breathing cycles are required to result in one deposited copy of RNA in the peripheral airways accompanied by 12 copy depositing in the extrathoracic, and one RNA depositing in the bronchial airways. Taking into account the maximal measured concentration of $1.34 \times 10^{11}$ RNA copy per cm$^3$, it can be obtained that from a single inhalation 4900 copy of RNA deposit in the extrathoracic airways, and 460 copy of RNA deposit in the peripheral airways. Considering that the concentration of viruses capable of infection is about one millionth of the RNA copy concentration (5), the probability of direct infection of the peripheral airways from inhalation of cough-generated aerosol is very low.

## 4    Discussion

It seems to be surprising that inhalation of aerosol generated by coughing of someone with viral loads of $4.69 \times 10^4$ RNA copy per g results in negligible amount of RNA copies, around $2.5 \times 10^{-4}$ in the peripheral airways of a bystander person. The reason for this fact is that the mass of inhaled material is low, and therefore the absolute number of pathogens is also low. Figure 5 shows the average number of RNA copies per aerosol particle as well as the probability that a particle contains at least one RNA copy of the virus as the function of particle size.

We assumed homogenous virus distribution in the initial throat swab sample, and that the number of viruses in a particle is proportional to its mass. The probability that a particle contains at least one RNA copy of the virus was calculated assuming Poisson distribution. The graph clearly shows that the probability that a particle contains a virus is relatively low in the submicrometer size range, even at very high virus concentrations in the throat.

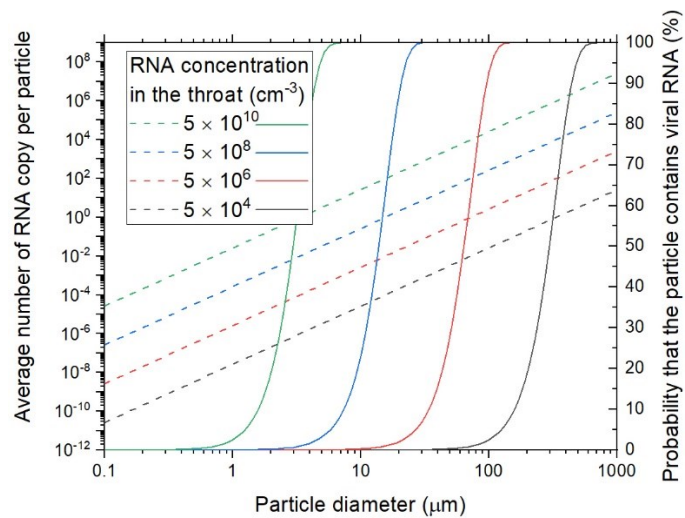

Figure 5. The average number of RNA copies of the virus in one particle (dash lines) and the probability that a particle contains at least one virus copy (solid lines) as the function of particle size in case of different initial virus concentrations in the throat (different colours).

Independently on the virus concentration in the inhaled air, it can be concluded that about 10 times more pathogens deposit in the extrathoracic airways than in the peripheral airways. It is in agreement with the clinical observation that the typical first respiratory symptom of COVID-19 is dry cough (22,23), which is often caused by upper respiratory infections. In addition, anosmia and ageusia have also been reported suggesting infection of cells in the upper airways (5,24,25).

It was also found here that material emitted by coughing directly reaches the peripheral airways. Therefore, inhalation of air after a bystander cough can theoretically induce pneumonia without the need for preceding upper respiratory symptoms. Based on the present study, however, such direct induction of COVID-19 pneumonia requires either very large virus concentrations in the aerosol particles or very long exposures to coughing. The latter underlines that even at home it is very important to cough into a tissue or cloth in order to absorb the emitted aerosol, and avoid the re-inhalation of own cough. Reducing the re-inhalation could then significantly prolong, or even block the onset of further, more severe phases of the COVID-19 disease.

The results suggest that inhalation of cough-generated aerosol containing SARS-CoV-2 results in only an upper airway infection directly, which can later develop into pneumonia. It is in agreement with the clinical observation that some patients who have mild symptoms initially will subsequently have precipitous clinical deterioration that occurs approximately one week after symptom onset (22,23,26). It is also in agreement with a recent virological analysis of nine mild cases of COVID-19 providing proof of active replication of the SARS-CoV-2 virus in tissues of the upper respiratory tract (5). They found very high concentration of viral RNA in and isolated the virus itself from early throat swabs. Our results suggest that without the enhancement of virus concentrations in the upper airways SARS-CoV-2 would be much less dangerous. In another study, nasal epithelial cells were found to show the highest expression of ACE2 among all investigated cell types in the respiratory system (9), which together with the deposition fraction synergistically increase the probability of upper respiratory infections.

From the throat where high virus concentration can be found in patients with mild symptoms (5), pathogens can be transported either via the cardiovascular or the respiratory system. As SARS-CoV-2 genome could not be detected in the blood of patients with mild symptoms (27), transport via the respiratory system can be the dominant route. Potential mechanisms of virus transport from the throat to the lower airways may include re-inhalation of own cough, aerosol generation in the throat during inhalation by resuspension, pathogen transport on the surface of the bronchial airways, or gradual infection of neighboring cells expressing ACE2 towards the periphery. Mucociliary clearance can inhibit the latter two processes suggesting that compromised mucus production or transport can be a risk factor for COVID-19 pneumonia. Understanding the processes leading from infections of the upper airways to pneumonia in COVID-19 patients may help to identify early treatment and prevention strategies, and provide further insights on the long incubation period.

## 5    Conclusions

In order to quantify the deposition distribution of cough-generated aerosol containing SARS-CoV-2 viruses, we applied the Stochastic Lung Deposition Model. It was found here that the probability of direct infection of the peripheral airways due to inhalation of aerosol by a bystander cough is very low. As the number of pathogens deposited in the extrathoracic airways is 10 times higher than in the peripheral airways, we concluded that in most cases COVID-19 pneumonia must be preceded by SARS-CoV-2 infection of the upper airways. The one week difference observed in several patients between the onset of their initial mild symptoms and precipitous clinical deterioration (22,23,26) provides a precious window for prevention of pneumonia and ARDS by blocking or significantly reducing the transport of the virus towards the peripheral airways. Therefore, coughing into a tissue or cloth even at home in order to absorb the emitted aerosol is highly recommended to avoid the continuous re-inhalation of own cough. Further research is required to understand the processes leading from infections of the upper airways to pneumonia in COVID-19 patients.

# 6    Conflict of Interest

The authors declare that the research was conducted in the absence of any commercial or financial relationships that could be construed as a potential conflict of interest.

# 7    Funding

The authors received no funding for this research.

# 8    Acknowledgments

The authors thank Tibor Kerényi, 2nd Department of Pathology, Semmelweis University, Budapest, Hungary, and Lajos Kovács and Dóra Krikovszky, 1st Department of Paediatrics, Semmelweis University, Budapest, Hungary for the helpful discussions.